\documentclass[english,aps,prb,onecolumn]{revtex4-1}
\usepackage{verbatim}
\usepackage{amstext}
\usepackage{graphicx}
\makeatletter
\draft 
\usepackage{subfigure}
\usepackage{multirow}
\usepackage{rotating}

\newcommand{\angstrom}{\mbox{\normalfont\AA}}

\makeatother

\usepackage{babel}
\begin{document}

\title{The thermodynamics of hydride precipitation: the importance of entropy, enthalpy and disorder}

\author{S. C. Lumley}
\affiliation{Department of Materials, Imperial College London, South Kensington, London, SW7 2AZ, UK.}
\affiliation{Nuclear Department, Defence Academy, HMS Sultan, Gosport, Hampshire, PO12 3BY, UK.}
\author{S. T. Murphy}
\affiliation{Department of Materials, Imperial College London, South Kensington, London, SW7 2AZ, UK.}
\author{P. A. Burr}
\affiliation{Institute of Materials Engineering, Australian Nuclear Science and Technology Organisation, Menai, New South Wales 2234, Australia.}
\affiliation{Department of Materials, Imperial College London, South Kensington, London, SW7 2AZ, UK.}
\author{A. Chroneos}
\affiliation{Engineering and Innovation, The Open University, Milton Keynes MK7 6AA, UK.}
\affiliation{Department of Materials, Imperial College London, South Kensington, London, SW7 2AZ, UK.}
\author{P. R. Chard-Tuckey}
\affiliation{Nuclear Department, Defence Academy, HMS Sultan, Gosport, Hampshire, PO12 3BY, UK.}
\author{R. W. Grimes}
\affiliation{Department of Materials, Imperial College London, South Kensington, London, SW7 2AZ, UK.}
\author{M. R. Wenman}
\affiliation{Department of Materials, Imperial College London, South Kensington, London, SW7 2AZ, UK.}
\email{m.wenman@ic.ac.uk}

\selectlanguage{english}%

\date{\today}
\begin{abstract}
The thermodynamics of H/$\alpha$-Zr solid solution and zirconium hydride phases were studied using density functional theory. Disorder in $\zeta$, $\gamma$ and $\delta$ hydrides and solid solutions were modelled using a statistically significant number of randomly generated structures in combination with special quasi-random structures and solid solutions with a range of concentrations. This is used in conjunction with a calculation of thermodynamic parameters of the system, including the temperature dependent sensible enthalpy, configurational entropy and vibrational entropy of the different crystals in the system, developed from phonon density of states. It was found that precipitation of hydrides is not thermodynamically favourable for Zr-H solid solutions containing less than 300 ppm H, suggesting that a mechanism must cause local concentration of H atoms to a greater amount than found globally in experimental samples containing hydrides. Temperature drives the reaction in the direction of solution, primarily due to entropic effects. Generally, $\gamma$ hydride is the most stable phase, although it is very close in energy to the $\delta$-phase. The sensible enthalpy of precipitation assists in stabilising HCP hydrides, and the configurational entropy change during precipitation favours FCC hydrides. None of the thermodynamic contributions are found to be negligible in driving precipitation.
\end{abstract}
\maketitle

\section{Introduction}

H in metals has been of considerable interest for many decades due to possible uses of metals as H storage devices but often more commonly because of the deleterious effect H has on metals through processes such as H embrittlement. There have also been numerous studies both by experiment and by first principles methods of the formation energies and stability of phases especially the precipitation of hydrides \cite{Smithson2002,Heuser2008,Tao2011,Nicholson2012}. One important system subject to hydriding is that of Zr alloys, often used for fuel cladding in light water nuclear reactors, due to their good mechanical and corrosion properties and low capture cross-section for thermal neutrons. The formation of hydrides in Zr alloys is a key issue in the ongoing development of fuel cladding for water cooled reactors \cite{Puls2012}. The rate of production, absorption, and precipitation of hydrides is one of the most important safety factors, as it is a limiting factor determining how long fuel assemblies may be kept in a reactor; it can cause complications with intermediary spent fuel storage \cite{Raynauld2011}.

Zr corrodes in an aqueous environment, producing an oxide layer and H. Once produced, some H is released into the water, and some diffuses through the protective oxide barrier, and into the Zr metal.

At reactor operating temperatures, H is soluble up to around 100\enskip ppm and extremely mobile in $\alpha$-Zr \cite{Kearns1967,Kearns1972, Barrow2013}. However, if sufficient H is present or if the solubility limit of H in $\alpha$-Zr is lowered (commonly due to a temperature drop during reactor transients \cite{Ells1968}), then H will precipitate out, forming a hydride. The hydrides are brittle and liable to cause failure of the fuel cladding; this is especially likely if they align along the radial direction of the fuel pin \cite{Puls2012,Mallett1957}.

Current understanding suggests that there are five main ways that H can be sequestrated in Zr metal \cite{Okamoto2006b,Weatherly1981,Zhao2008,Domain2002}; these are, solid solution and four hydrides: $\zeta$, $\gamma$, $\delta$ and $\varepsilon$. The structures of these different hydrides are shown in Fig \ref{fig:Hydrides}. There is a general relationship of decreasing c/a ratio with increasing H content. Of particular interest is that the most commonly observed $\delta$ phase is often reported to have a formula of ZrH$_{1.66}$ and a disordered fluorite structure, while simulators often use a simplified ersatz of periodic ZrH$_{1.5}$ \cite{Zhu2010,Domain2002}. The $\gamma$ hydride has been assumed to be meta-stable, as it is less readily observed than the  $\delta$ hydride \cite{Lanzani2004}. However, other investigations have observed the $\gamma$-hydride at room temperature conditions \cite{Root1996}, under both slow cooling \cite{Weatherly1981} and fast cooling regimes \cite{Nath1975}. Overall, it appears that the stability and occurrence of these phases is a complex phenomenon, where the H concentration, thermal treatments, alloying additions and stress states all have a part to play in determining which phases are observed \cite{Steuwer2009}.

The focus of this study is on modelling hydrides using atomistic simulation techniques based on density functional theory (DFT). DFT has previously been used to successfully model thermodynamics in H-metal systems \cite{Heuser2008,Tao2011}. In this system in particular, it has been found that H preferentially occupies tetrahedral sites in the Zr lattice, rather than other sites \cite{Domain2002,Burr2012,Fukai1993}. Some works have focused on the octahedral site as the main location for H atom solution, however such works appear to be in the minority \cite{Glazoff2013,Glazoff2014}. It has been theoretically predicted that the FCT structure of ZrH$_{2}$ can have another stable phase for c/a > 1 \cite{Ackland1998,Blomqvist2012}. Zhu et al. \cite{Zhu2010} studied the ordered hydride phases using DFT; they concluded that $\delta$-ZrH$_{1.5}$ is thermodynamically less stable than the other phases at high temperature. Zhong and MacDonald, who published previous DFT results combined with new calculations of the $\gamma$ phase, suggest the $\gamma$ phase is stable at temperatures below about 523 K\cite{Zhong2012}.

The simulation of  hydrides is complicated due to the random distribution of the H atoms in some phases. Although hydrides have been simulated in the past, few studies have attempted to examine hydrides whilst taking into account the disorder. One of the most successfully used techniques for simulating disordered atomic structures is the special quasi-random structures (SQS) method developed by Zunger et al. \cite{Zunger1990}, which has been used to simulate a range of non-stoichiometric materials and structures. It has also recently been applied to this system \cite{Wang2013}. In this study SQS techniques are combined with a statistical analysis of a large number of randomly generated cells, in order to examine the impact of disorder on hydride precipitation. Phonon calculations are also used to calculate thermodynamic properties such as the vibrational entropy and the sensible enthalpy changes during precipitation. Previous studies on other systems by DFT have shown the importance of vibrational entropy on the solubility limit in precipitation reactions from solid solution, and its importance in creating a temperature dependent understanding of hydride precipitation \cite{Akbarzadeh2009,Mao2013}. Thus, a comprehensive view of the enthalpic and entropic contributions towards hydride precipitation in Zr is developed.

\section{Methodology}

\subsection{Simulation Parameters}
For this investigation, CASTEP 5.5 was used to simulate the different structures \cite{CASTEP}. As a plane-wave pseudopotential code, it is particularly appropriate for modelling crystals.  Ultrasoft pseudopotentials were generated "on-the-fly", under the formalisation of Vanderbilt et al. \cite{Vanderbilt1990}.Valence electrons for Zr were modelled as $4s^{2}4p^{6}5s^{2}4d^{2}$. Convergence with respect to basis-set cut-off energy and k-point grid density was tested in a series of electronic self consistency calculations. It was found that the simulations were accurate to 2 d.p.\ for a cut-off energy of 400 eV and a k-point grid spaced at 0.03 $\angstrom^{-1}$. k-points were arranged in a gamma-centred Monkhorst-Pack grid \cite{Monkhorst1976}. As the system displays metallic characteristics, the integration of the Brillouin zone is achieved via a Methfessel-Paxton scheme , with a band smearing width of 1 eV\cite{Methfessel1989}.

All cells used in this work are geometry relaxed in order to approach their minimum energy configuration. Cells were considered relaxed when the difference between two successively modified iterations were below all of the following criteria:

\begin{itemize}
\item Energy derivative $<$ 0.001 eV
\item Force on ions $<$ 0.05 $\textrm{eV\angstrom}^{-1}$
\item Displacement of ions derivative $<$ 0.001 $\textrm{\angstrom}$
\item Total stress derivative $<$ 0.1 GPa
\end{itemize}

Relaxation of atomic positions was carried out under the quasi-Newtonian BFGS scheme \cite{Pfrommer1997}. Both atomic positions within the cell, the lattice constants and cell aspect ratios were unconstrained during relaxation. This means that volume and cell distortions due to H accommodation are fully accounted for.

\subsection{Thermodynamic Considerations}
With static simulations it is usual to consider the energy change from a set of reactants to a set of products. Relating the calculated energy changes to a real system is often difficult, as there are numerous different energy components, while static calculations can directly only evaluate the ground state changes (ie. the latent enthalpy of a reaction). Thus, many factors present in a real system are not represented in a simulation. These different components and factors are often poorly communicated between experimentalists and simulators, leading to inappropriate assumptions when using terms such as energy, enthalpy and entropy. To ensure a clear understanding, the terms that were calculated here are now discussed.

The fundamental measure of the driving force behind a reaction is the Gibbs free energy change:

\begin{equation}
\Delta G = \Delta H - T\Delta S
\label{eq:gibbs}
\end{equation}
where $\Delta H$ represents the enthalpy change of the system and $\Delta S$ represents the entropy change. There are different contributors to the enthalpic and entropic terms, thus Eq. \ref{eq:gibbs} can be expanded into:

\begin{equation}
\Delta G = (\Delta H_{l}+ \Delta H_{s T}) - T (\Delta S_{v} + \Delta S_{cl})
\label{eq:gibbs2}
\end{equation}
where:

\begin{itemize}
\item $\Delta H_{l}$ is the latent enthalpy associated with the reaction. It is due simply to the formation or destruction of bonds and is independent of external conditions. When quoting results from DFT simulations, this is the most commonly reported number. If the other terms are ignored this is often quoted as the energy.

\item $\Delta H_{s}$ is the sensible enthalpy and is related to the heat capacities of the different species involved in the reaction. In order to calculate this, the amount of thermal energy that can be stored in the lattice needs to be known. This is achieved by calculating the phonon density of states of the materials in question, and integrating the different acoustic modes in a quasi-harmonic approximation, for a temperature $T$, to yield $H_{s T}$.

\item $\Delta S_{v}$ is the vibrational entropy and is a function of the number of discrete vibrational energy levels that exist amongst the different acoustic modes of the system. This is determined by the application of the Boltzmann entropy equation to a quasi-harmonic approximation of the phonon distribution\cite{Baroni2001}.

\item $\Delta S_{cl}$ is the lattice component of the configurational entropy. This only applies to solids with a disordered lattice, which in this system consists of the H sub-lattice. An approximation of this is reported in section \ref{sec:Entropy}.

\end{itemize}

It is clear that the distribution of phonons must be calculated in order to determine two of the terms included here. Analysis of phonon distributions has been previously examined in this system by Blomqvist et al. \cite{Blomqvist2012}. In the present study, this was achieved by means of the finite displacement method, in the (direct) supercell approach \cite{Frank1995}. The application of the phonon data to the creation of thermodynamic parameters has been described previously \cite{Baroni2001}.

In a real precipitation event, there would be stress related effects, as well as contributions arising from the creation of a hydride-Zr interface. There would also be entropy created by intrinsic defects on the Zr lattice. This study cannot comment on such effects, as they would require simulations that are either significantly larger than the scale that DFT presently allows, or an exceedingly large number of simulations. However, the numbers produced in this work represent a large portion of the overall driving forces for precipitation, and are a significant advance on what has been achieved for this system previously.

\subsection{Cell Configurations}

This study examines both ordered and disordered models of hydrides and Zr-H solid solutions. The ordered models are performed by a straight-forward geometry optimisation of hydride structures including the commonly accepted structures of $\zeta$-Zr$_{2}$H, $\gamma$-ZrH, $\delta$-ZrH$_{1.5}$ and $\varepsilon$-ZrH$_{2}$ \cite{Domain2002, Zhao2008}. In addition, dilute solid solutions were also simulated in which one hydrogen atom was placed in $2\times2\times2$, $3\times3\times2$ and $4\times4\times3$ supercells of $\alpha$-Zr, giving concentrations of 5.8 at\%H, 2.7 at\%H and 1.0 at\%H respectively. A further calculation was carried out in which one atom of H was placed in a $2\times2\times2$ supercell of FCC Zr. All of the solid solution calculations described so far were carried out with H occupancy being investigated on both octahedral and tetrahedral sites.

A different method was used to simulate more disordered structures. As a starting point for the generation of off-stoichiometric hydride phases, a $\delta$ hydride structure is built, formed from a $2\times2\times2$ supercell of the primitive cell, with all tetrahedral H sites occupied (giving a formula of ZrH$_{2}$). This is similar to the $\varepsilon$ hydride, except that the $c/a$ ratio is 1. A new cell is generated from this input, by giving each H atom a random chance to be removed of 0.166 (or $1-\frac{1.\dot{6}}{2}$). This is repeated to generate a large number of cells, with the only constraint being that new cells must be unique from previously generated cells. These cells then undergo geometry optimisation to provide energies for each configuration. No constraints are placed upon the number of H atoms in any given cell, meaning some cells will have more H than the 1.66 ratio, and some will have less. Providing the number of cells is large enough, the formulae of the different cells will follow a normal distribution with a mean of 1.66. A similar method was used with a $2\times2\times2$ supercell of $\alpha$-Zr, and a removal probability of 0.89 with single H atom cells discounted. This provides a selection of solid solution cells where the H exists in small clusters, modally containing 2 H atoms, which are used to bridge the stoichiometry gap between a dilute solid solution and the $\zeta$-phase. These sets will have a number of different possible configurations, which allows examination of how configurationally sensitive the properties of the hydrides and solid solutions are. Taken together, an arbitrarily large set of these cells provide a more complete description of the $\delta$-hydride and a concentrated solid solution than any single periodic calculation. The advantages of random structure generation have been discussed previously \cite{Pickard2011}.

In order to qualify this method against established techniques, the random $\delta$-hydride set is compared with SQS generated cells. The SQS technique, as detailed in reference \cite{Zunger1990}, works on the assumption that in any random arrangement of atoms on a pre-defined set of atomic sites, some clusters of atoms will be more common than others. The more common clusters are defined by the structure and stoichiometry of the simulated crystal. Thus, a number of "special" cells can be constructed that comprise the more common configurations, and less common configurations can be discounted. In this study, 3 different SQS cells are simulated containing 48, 52 and 56 H atoms, along with 32 Zr atoms. The SQS method has been recently applied to this system in order to model bulk parameters \cite{Wang2013}. The 3 most representative SQS cells in that paper are used again here.

Some phases, however, exhibit more order than the $\delta$ phase, such as the $\gamma$ and $\zeta$ phases. Generally, these phases are simulated as ordered phases with a defined structure and no disorder. However, any real hydride will likely exhibit a degree of disorder, particularly when a disordered phase can transform to an ordered phase and vice versa. Thus, a method has been devised to introduce disorder into these structures, whilst biasing the results towards the known structures. In this "skew-random" method, the starting point is a cell of the ordered structure. For each occupied H site, a small removal probability of the H atom is introduced. Likewise, for each unoccupied site, a small chance is created to have a H atom inserted. In both of these cases the probability of 0.05 was used, as it created a reasonable spread of random structures, whilst remaining close to the crystallographic form of the hydride. As before, a large number of these cells were generated and only unique cells are simulated to ensure that the full ensemble of simulations is relevant to a (partially) disordered hydride. A smaller number of random cells was also generated with no structure biasing and a probability for removal of 0.5. These were used purely as a comparison with the skew-random cells.

Ultimately, all possible arrangements of H on interstitial sites in Zr exist as points in the configuration space. Each of these different methods develops a different sampling of that configuration space, aiming to ensure that a valid distribution is identified. Such sampling methods are required, as a single point is not sufficient to model all the important aspects of the system, while a full census of the configuration space carries a computationally prohibitive cost.

\section{Results}
\subsection{Elements and Ordered Crystals}
In order to ensure the validity of the simulations performed, it is important to confirm that the simulation results can reproduce experimental data. As much of the more useful thermodynamic data makes use of the pure elements as reference states, it is particularly important to ensure that the reference states are well modelled. In Table \ref{tab:elements}, properties of reference states are reported compared with established literature values. Excellent agreement is achieved on all counts, to within a maximum discrepancy of 1.65\%. As most of the thermodynamic values in this study are contingent on the presence and position of H atoms in a Zr cell, the values obtained for H are of critical importance. For this reason, great care has been taken to ensure that the simulation parameters can accurately produce the properties of the H$_{2}$ gas reference state.

The enthalpies of formation of different ordered crystals are given in Table \ref{tab:ordered}. The formation energies were calculated via the equation:

\begin{equation}
\Delta E^{F} = \frac{1}{x+y}\left[E(\mathrm{Zr}_{x}\mathrm{H}_{y}) - \left(E(x\mathrm{Zr}) + y\frac{1}{2} E(\mathrm{H}_{2})\right)\right]
\label{eq:Formation Enthalpy}
\end{equation}
where $E(Zr_{x}H_{y})$ represents the energy of the hydride (or solid solution containing H) and the other terms refer to the pure elements. As the number of atoms in each simulation differs, the formation energy must be normalised with respect to the number of atoms, in order to not bias the formation energies towards the larger cells. Solution enthalpies of H atoms in a HCP and an FCC matrix are also presented. Although the latter configuration is un-physical (pure FCC Zr is not a stable phase, nor observed in real alloys) it does provide useful comparison points. Although this phase assumes the absorption of H into an FCC lattice, the reference state is still taken to be HCP Zr. This is done to ensure fair comparison with other results and is based on the assumption that any starting point that could lead to this configuration would still be based on HCP Zr. As formation energies for FCC solutions must also contain the energy associated with a HCP $\rightarrow$ FCC phase change, it is reasonable that the formation energies for the FCC solutions are higher than their HCP counterparts. The dilute tetrahedral solid solution is of particular importance as it represents a reference point for comparison in further calculations. The number reported here of -0.60 eV compares favourably with -0.52 eV from \cite{Udagawa2010}, -0.604 eV from \cite{Domain2002} and -0.464 from \cite{Burr2013}. The tetrahedral site for H occupancy remains the most favourable, in agreement with most prior work \cite{Udagawa2010,Burr2012,Domain2002,Burr2013}. For the remainder of this work, when considering sites for H occupancy, only the tetrahedral site is considered. More exotic configurations such as H$_{2}$ dimers on interstitial sites have also previously been found to be unfavourable \cite{Domain2002}.

With regards to the approximate $\delta$ phase, ZrH$_{1.5}$, a conventional unit cell of FCC Zr offers 8 sites for H occupancy, 6 of which must be filled with the other two vacant. If the system is cubic, then symmetry reduces the number of configurations to three different arrangements. These arrangements are where the vacancies are both in the [100], [110] and [111] directions. These are referred to in Table \ref{tab:ordered} by these directions. The configuration with the lowest enthalpy of formation is the one in the [111] orientation, where the vacancies are separated by the longest distance. The energy difference between these states is relatively small, and similar (but slightly larger) than the average of the energy calculated from using the three SQS configurations.

\subsection{Statistical Analysis}
In order to ensure that the simulations are representative of the disordered system it is important that a large enough sample of the configuration space is achieved. With this in mind, the statistical parameters generated in the sets used are shown in Table \ref{tab:Stats}. The $\delta$ and solid solution series rely on a random distribution about the selected stoichiometry, while the $\zeta$ and $\gamma$ phases use the skew-random method. A large enough sample has been made when the data set forms a normal distribution centred on the target stoichiometry. A simple convention for determining normality is a plot of the cumulative distribution probabilities of the data against theoretical cumulative distribution probabilities generated by a standard normal distribution \cite{Chambers1983}, as created from the parameters in Table \ref{tab:Stats}. A straight line fit would represent perfectly normal data. Normality tests were performed on sets of increasing sample size until a high degree of confidence in normality was achieved. Figs \ref{fig:Statform} and \ref{fig:Statenth} show normality tests for each set of data, generated from both the stoichiometry distribution and the formation enthalpy distribution. In both cases, we see all series display a good linear fit. We report that for sample sizes of 50 cells per set, all datasets showed high coefficients of linear regression with the lowest R$^{2}$ being 0.9602. The average stoichiometry for each set is extremely close to the experimental formula value considered representative for that hydride structure. This gives confidence to the hypothesis that this set of randomly generated structures approximates a disordered material when taken as a whole.

\subsection{Enthalpies\label{sec:Enthalpy}}
Relative thermodynamic phase stability can be determined by plotting the formation energy across the range of compositions. These results are presented in Fig. \ref{fig:FormationEnthaply}. Specifically, this represents a latent formation enthalpy, as opposed to a free energy. In addition to a presentation of the various data sets, a convex hull has been drawn on the plot. Here, a convex hull is defined as the smallest convex path to contain all of the available data points, when viewed from below the plot. It is useful, because any mixture with an enthalpy less negative than the convex hull would be more stable as a mixture of the two configurations which bound that segment of the hull.

In Fig. \ref{fig:FormationEnthaply}, all enthalpies are negative indicating that there is a general thermodynamic driving force for formation, which becomes stronger with greater H-content phases. However, the majority of configurations lie above the convex hull, indicating they are less stable than a mixture of other phases. The configurations which lie on the convex hull are the 1 at\%H solid solution, the stoichiometric and ordered $\gamma$-hydride, and the $\varepsilon$-phase with a c/a ratio of less than 1. The enthalpies in the stoichiometry range ZrH$_{0}$ to ZrH$_{0.6}$ agree with a similar plot produced by Hollinger et al. in terms of the range of formation enthalpies of the different structures \cite{Holliger2009}. However, whereas that work noted stable structures in this range, none are found in the present work. This is almost certainly due to the fact that the work of Hollinger et al. was focused on hexagonal phases, whereas the present work shows that cubic phases out-compete hexagonal structures in terms of stability. Domain et al. \cite{Domain2002}, provide a similar plot with no convex hull, however adding one demonstrates the same phases ($\varepsilon$ and $\gamma$) as stable and by similar energies. This result has also been found by Zhong and MacDonald, who used this data to suggest that the $\gamma$ phase is thermodynamically stable below $\approx$ 523 K \cite{Zhong2012}. This contrasts greatly with Zhu et al. \cite{Zhu2010}, who claimed that the $\delta$-hydride is by far the most stable hydride, by nearly 8 eV more than the other phases. It is, however, difficult to understand that result, since the "convex hull" presented was not actually convex, and the magnitude of this number is out of line with other results \cite{Domain2002,Holliger2009,Udagawa2010,Zhong2012}.

Ultimately, hydrides are formed by the precipitation of H from solid solution in the $\alpha$-Zr matrix. This reaction is given by the expression:

\begin{equation}
\Delta E^{P} = \left[E(Zr_{x}H_{y})+(y)E(Zr_{R})\right] -  \left[yE(Zr_{R}H)+ xE(Zr)\right]
\label{eq:Precipitation Enthalpy}
\end{equation}
This equation forms the basis of calculating the change in different thermodynamic parameters such as the latent enthalpy of precipitation. The term $R$ is the number of Zr atoms in the solid solution reference cell. Precipitation enthalpies have been calculated using reference solutions containing 96, 36 or 16 atoms of Zr to one atom of H. This equation is balanced with free Zr on both sides because it ensures that the reaction maintains reversibility in situations where the concentration is different. As with the formation enthalpies, these precipitation enthalpies must be normalised to ensure that larger simulations are not shown as having larger enthalpies purely based on their size, and not on changes in composition and thermodynamic behaviour. To this end, all simulations are divided by the total number of H atoms present in the hydride phase, and then converted into kJ\enskip mol$^{-1}$. Thus, the enthalpies presented hence forth are in units of kJ\enskip molH$^{-1}$, representing the enthalpy change required for one mole of H atoms to precipitate from a solid solution.

The latent enthalpies of precipitation are presented in Fig. \ref{fig:LatentEnthaply}. As with the formation enthalpies, there is a general trend that the precipitation of H-rich hydrides is more preferable than H-poor hydrides. On the H poor side of the graph, solid solutions have more negative enthalpies when they are less concentrated than the reference solid solution for that series, suggesting a trend towards dilution of H atoms. Moving across towards products with a greater H content, there is then a peak of unfavourable H clusters around ZrH$_{0.2}$, followed by a steady return to the more preferable hydride phases. In particular, the $\gamma$-phase exhibits the strongest preference for precipitation, with the most favourable configuration being the structure typically modelled in other simulation studies, shown in Fig. \ref{fig:Hydrides}. There is a notable discontinuity in all series at $\approx$ ZrH$_{0.75}$, corresponding to the point where the series switched from modelling HCP hydrides to FCC hydrides.

There are no negative latent enthalpies of precipitation for the precipitations from 1 at\%H and 2.7 at\%H solid solutions. However there are for the 5.9 at\%H solid solution. It is sensible that increasing the H content in the Zr lattice increases the impetus for the rearrangement of the H atoms into a hydride, as is evidenced by the existence of a terminal solubility limit for H in Zr \cite{Kearns1967}. Overall, this plot is consistent with H having a bimodal distribution in Zr, preferring to exist either as a sparsely distributed solid solution, or as a concentrated hydride. A middle-ground between these two modes is unfavourable.

So far, this only describes the latent enthalpy with no regards for the effects of temperature. The sensible enthalpy of precipitation is related to the heat capacities of the products and reactants of the precipitation reaction. Heat capacities calculated at 298 K are 23.00 kJ\enskip mol$^{-1}$ \enskip K$^{-1}$ for $\alpha$-Zr and 28.64 kJ\enskip mol$^{-1}$ \enskip K$^{-1}$ for $\varepsilon$ ZrH$_{2}$ (compared with the available experimental values of 25.45 kJ\enskip mol$^{-1}$ \enskip K$^{-1}$ and 31.08 kJ\enskip mol$^{-1}$ \enskip K$^{-1}$ \cite{CRC1999}). Fig. \ref{fig:AbsSensibleEnthalpy} gives the absolute sensible enthalpies for reference simulations, as they vary with temperature. The values for the three $\delta$-phase stoichiometries are calculated using the SQS generated cells. There is a general trend for increasing sensible enthalpy with increasing H content. The $\zeta$-phase has a substantially lower sensible enthalpy than the other hydride phases. As the temperature increases, the variance in sensible enthalpies decreases. The enthalpy calculated at 0 K represents the zero point energy contribution to the enthalpy of this system.

The absolute sensible enthalpy is of less interest than the change in sensible enthalpy which may drive precipitation. Using Fig. \ref{fig:AbsSensibleEnthalpy}, a surface is generated to describe the relationship between composition, temperature and sensible enthalpy. Using this surface, values are interpolated for sensible enthalpies for all the cells examined in this work. The enthalpy data is sufficiently close that a simple linear interpolation does not introduce unreasonable variance. Feeding this interpolation into Eq. \ref{eq:Precipitation Enthalpy} (replacing the latent enthalpy term), the sensible enthalpy change during precipitation for a variety of different structures and temperatures is generated and plotted in Fig. \ref{fig:SensibleEnthalpy}. This information is presented only for the precipitation from the 16 atom solid solution. Given that the sensible enthalpy is added to the latent enthalpy, in Eq. (2), a negative value of sensible enthalpy represents a driving force for precipitation, while a positive value represents a driving force for solution. The sensible enthalpy appears to drive the system towards precipitation for all product stoichiometries greater than $\approx$ ZrH$_{0.08}$. As temperature increases, the driving force for precipitation also increases. There is a relative increase in this driving force for stoichiometries of $\approx$ ZrH$_{0.4}$, which corresponds roughly with the stoichiometries found in the $\zeta$-phase hydrides. The sensible enthalpy then becomes less negative for stoichiometries appropriate to $\gamma$-hydrides before reducing slightly for hydrides with even greater H content.

\subsection{Entropy \label{sec:Entropy}}
As described previously, computing the free energy of a reaction requires a description of the entropy as well as the enthalpy. In this study, we examine two sources of entropy - the vibrational and the configurational.

Configurational entropy stems from the disorder available when the structure may have multiple different forms. It is quantified by the Boltzmann entropy equation:

\begin{equation}
S_{c} = k \ln\Omega
\label{eq:Boltzmann}
\end{equation}
where $\Omega$ is defined as the number of different configurations or micro-states in which the system may be arranged and $k$ is Boltzmann's constant. In an atomistic context, the number of different structure configurations is given by adapting the standard permutations expression:

\begin{equation}
\Omega= \frac{(N_{V}+N_{H})!}{N_{V}!N_{H}!}
\label{eq:Permutation}
\end{equation}
where $N_{V}$ is the number of potential H sites which are vacant, while $N_{H}$ is the number of H atoms.

As before, the primary concern is not the absolute entropy, but the change in entropy during precipitation. Using the entropy calculated in Eq. \ref{eq:Boltzmann} in the precipitation equation (4) (with the H coming from the 16 atom Zr cell), the change in configurational entropy is determined across a range of stoichiometries, and displayed in Fig. \ref{fig:ConfigEntropy}. These entropies are presented as a $T \Delta S$ product. As entropies are subtracted from enthalpies to generate a free energy, a negative value indicates a driving force towards solution while a positive value drives towards precipitation. For non-zero temperatures, we see that the configurational entropy represents a driving force for solution, with increasing temperatures, of course, leading to more negative values. There is a notable discontinuity when the simulated series shift to modelling FCC hydrides. This is because the FCC structure has more tetrahedral sites per Zr atom, which are considered as possible sites for H occupancy (i.e. it offers greater configurational options). Thus, the shift from HCP to FCC is favoured by the configurational entropy and this driving force increases with temperature.

The final contribution examined in this work is the vibrational entropy. Vibrational entropies are shown in Fig. \ref{fig:AbsVibrationalEntropy} for the same reference cells as used in calculating the sensible enthalpy. Vibrational entropies at 298 K are 37.52 J\enskip mol$^{-1}$ \enskip K$^{-1}$ for $\alpha$-Zr and 31.387 J\enskip mol$^{-1}$ \enskip K$^{-1}$ for $\varepsilon$ ZrH$_{2}$, compared with the available experimental values of 39.144 J\enskip mol$^{-1}$ \enskip K$^{-1}$ and 35.154 J\enskip mol$^{-1}$ \enskip K$^{-1}$ respectively \cite{CRC1999}. It should be noted that experimental results will include other forms of entropy (such as that generated by intrinsic defects on the Zr lattice), hence it is reasonable that the theoretical results are slightly smaller than experimental values. Fig. \ref{fig:AbsVibrationalEntropy} demonstrates a decreasing vibrational entropy with increasing H content. Similar to the calculation of sensible enthalpies, this plot is used to interpolate values from a temperature-composition-entropy surface. Applied across the range of compositions, the vibrational entropy is given in Fig. \ref{fig:VibrationalEntropy} as a T$\Delta$S product. This plot shows negative values for all compositions above $\approx$ ZrH$_{0.08}$, and temperatures above 0 K. This is consistent with the vibrational entropy driving the reaction towards solution, with the effect becoming stronger with increasing temperature. The vibrational entropy change during precipitation is positive for dilute solid solutions, becomes negative for non-dilute solutions, and becomes more negative as H content increases. There is a decrease in the magnitude of the entropy change for hydrides of around $\approx$ ZrH$_{1.5}$, suggesting vibrational entropy may contribute to stabilising the $\delta$ phase.

\subsection{Free Energy}
With the change in both the enthalpy and entropy terms calculated for the precipitation reaction, the overall free energy change can be calculated from Eq. \ref{eq:gibbs2}. It is sometimes stated that vibrational entropies and sensible enthalpies are too small to be important in this system and other hexagonal metals \cite{Nicholson2012}. Although this may be true for predicting if hydrides occur at all, and for determining energies when one reactant is in a different state (eg. H$_{2}$ gas), in a system with multiple phases, containing subtle interactions, the magnitude of these other terms may be important. Given that the sensible enthalpy, configurational entropy, and vibrational entropy are all within the range of -20 to 40 kJ\enskip molH$^{-1}$, none of these variables can be discounted and all have a part to play in determining phase stability.

In Figs. \ref{fig:FreeEnergy1} - \ref{fig:FreeEnergy3}, the lowest energy configuration from each data set is plotted as a free energy, with respect to the stoichiometry. Fig. \ref{fig:FreeEnergy1} represents the free energy of precipitation from the 96 atom cell, Fig. \ref{fig:FreeEnergy2}, is from the 36, and Fig. \ref{fig:FreeEnergy3} is from the 16. In the first of these plots, Fig. \ref{fig:FreeEnergy1}, the free energy remains positive across the entire stoichiometry range. Temperature raises the energy by over three times the 0 K values. As with all these plots, there appear to be five distinct regions, defined by stoichiometry. The first occurs between ZrH$_{0}$ $\rightarrow$ ZrH$_{0.1}$. Here, increasing the stoichiometry drastically increases the free energy of precipitation, suggesting that concentrating the H in the lattice is energetically unfavourable. This reaches a relatively flat region 2, made up of clusters of H atoms. This is particularly unfavourable, suggesting that H prefers to remain distributed. There is then a significant drop in free energy entering into region 3. The start of region 3 contains both H clusters, and sub-stoichiometric $\zeta$-hydrides as modelled by the skew-random technique. The $\zeta$-hydrides are more energetically favourable, and have a minimum energy point at ZrH$_{0.5}$, for the expected structure of the $\zeta$-hydrides. However, as  H content continues to increase, the free energy rises again and is out-competed by the sub-stoichiometric $\gamma$-phase at the start of region 4. This phase remains competitive until region 5 is entered, where stoichiometries are closer to that of the $\delta$-phase than $\gamma$. Beyond this, energies remain relatively flat until the terminating $\varepsilon$-phase is reached. As temperatures increase, region 5 begins to show a slight upwards slope, signifying it is more preferable to precipitate larger quantities of H-poor hydrides, than smaller quantities of H-rich hydrides.

The results are similar with an initial concentration of 2.7 at\%H in Fig. \ref{fig:FreeEnergy2}. Although the same region structure follows, the 1.0 at\%H solid solution, $\gamma$ and $\varepsilon$-phases are now negative, showing precipitation is now favourable at 0 K. At higher temperatures, the reaction is still driven towards solid solution. Thus, the kinetics of H transport will mean that H is maintained in isolated solid solution. The free energies are lower overall, and the difference brought about by increasing temperature is smaller.

Finally, the free energies drop significantly when moving to an initial concentration of 5.9 at\%H. Fig. \ref{fig:FreeEnergy3}, shows negative precipitation energies across the full range of stoichiometry and temperatures, with the exception of stoichiometries in the range of ZrH$_{0.1}$ to ZrH$_{0.4}$ (region 2). Below this, there is still a thermodynamic driving force for keeping H dispersed, but above this, there is impetus for hydride formation. As in all prior plots, increasing temperature cause free energies to become more positive and drives solution. At the higher temperatures, the energy of the $\zeta$-phase increases to the point where its precipitation is no longer thermodynamically favourable. The tendency for regions 4 and 5 to slope upwards at higher temperatures is even greater when precipitation occurs from a more concentrated initial solution. The most favourable phases are $\gamma$ and $\varepsilon$, although higher temperatures seem to favour $\gamma$-hydrides. Overall, this remains consistent with a bimodal H distribution, as described previously.

\section{Discussion}

\subsection{Hydrogen in zirconium}
It is well established that there is a strong thermodynamic impetus for H to become sequestrated in a Zr lattice. This is corroborated by the energy values given for H solution in Table \ref{tab:ordered}, which are all substantially negative. If we compare this for any of the values for precipitation, we see that the impetus of adding H to the Zr lattice is substantially greater than the energy of rearranging or precipitating the H once it is already in the lattice. This means that if the thermodynamic values for H$_2$ gas are less affected by temperature than the solid solution values are, it would be expected that H will continue to be added to Zr over the life of the cladding, steadily driving up concentration.

In Figs. \ref{fig:FreeEnergy1} and \ref{fig:FreeEnergy2}, the lowest free-energy configuration is the 1.0 at\%H solid solution, suggesting that H will preferentially form a dilute solid solution if possible. Temperature effects drive this behaviour further, in that the energy of the 1.0 at\%H solid solution becomes more negative and the more H rich solid solutions become more positive. In order to produce this behaviour, there must be some sort of interaction between H atoms that raises the energy of the system. Given that electron interactions have been previously demonstrated to be extremely localised to H atoms in the H-Zr system \cite{Domain2002}, it is unlikely that the chemistry of H is driving this response. This leaves geometrical factors and most notably stress. It is possible that the stress fields created by the insertion of H atoms into nearby interstitial positions in the Zr lattice are mutually repulsive.

\subsection{Implications for hydride precipitation}
If there is an initial impetus for H atoms to remain in solid solution, then given that hydrides have been noted to form experimentally, at some point conditions \textit{must} change to favour hydride formation. As more and more H atoms are absorbed by the Zr, the barrier for H atoms to congregate must be overcome. In Fig. \ref{fig:FreeEnergy3}, a high starting concentration of 5.9 at\%H provides this condition. This suggests at some point between 307 ppm and 690 ppm, that H atoms will be so numerous that they will be pushed past their mutually repulsive behaviour and will start to form hydrides. These values are significantly higher than those measured globally in actual alloys, implying Zr has a much higher local H carrying capacity than suggested experimentally \cite{McRae2009}. If this is the case, in order for precipitation to occur, forces beyond those predicted in these DFT simulations must generate a driving force for concentration, and hence precipitation. Given this case, it is possible that larger scale stress states, such as those provided by defects or cracks, could lessen this mutually repulsive force, allowing H atoms to diffuse together more easily. The idea that stress impacts diffusion is not new, and has been suggested as a key part of the mechanism behind DHC \cite{Puls2012,Puls2009,McRae2009}. It is reasonable that the Zr lattice around an interstitial H atom is in a state of compression. Given this, a tensile stress field would provide a nullifying effect on the repulsive interaction. Coupled with areas that are not under stress, there would be an impetus for H atoms to move away from regions where H atoms are in close proximity and not under tension, towards areas where they can congregate more favourably. This argument is based upon the existence of the aforementioned barrier to association, and provides an area which can be investigated in future studies.

Temperature also has the effect of driving the system towards solution, by raising the composition of the first point where hydriding may occur. In the room temperature series (300 K), the first hydride with a negative free energy of precipitation has a composition of about ZrH$_{0.43}$, while in the operating temperature series, this is raised to over ZrH$_{0.7}$. The main reason for this increase is entropic, in that both the configurational and vibrational entropy drive the reaction towards solution.

\subsection{Zirconium Hydrides}
The free energy curves produced in this study can be used to predict which hydride phase will precipitate, should precipitation occur. At 0 K, the most stable hydrides in all free energy figures are either the $\gamma$ or $\varepsilon$ phases. The commonly found $\delta$-phase rests somewhere between these two structures. Fig. \ref{fig:FreeEnergy3} suggests that as temperature increases, the stability of the $\gamma$-phase relative to $\varepsilon$ increases. The primary driver of this is the configurational entropy. Thus, in reactor, we would expect to see $\gamma$-phase hydrides, while at room temperature or lower, hydrides with a greater H content are favoured. Due to the difficulties present in observing hydrides during operation, the exact morphology of hydrides present in a reactor are unknown and are generally inferred from low temperature experiments. However, it has been noted that $\gamma$-phase hydrides are present in both slowly and rapidly cooled samples, suggesting that the $\gamma$-phase can remain stable under low temperature equilibrium conditions \cite{Weatherly1981,Ells1968}. There are, however, other factors in the precipitation of hydrides that are difficult to investigate using DFT. For example, although the work here provides the thermodynamic driving force for precipitation of a quantity of hydride from a quantity of Zr-H solid solution, there is also the issue of forming an interface, which may or may not be coherent, resulting in  additional parameters. If such parameters increased the energy of the $\gamma$ or $\varepsilon$-phases, then the $\delta$-phase could be more preferable.

The recently found $\zeta$ phase is normally considered to be a metastable phase \cite{Zhao2008}. The precipitation energies calculated here support this view, as the free energy curve forms a local minima about the ZrH$_{0.5}$ composition. The sensible enthalpy plays a large part in the stabilisation of this phase over others. Interestingly, it is the only phase in Fig. \ref{fig:FreeEnergy3} that shows a transition from negative to positive precipitation energy with increasing temperature. The results suggest the $\zeta$ phase should not be observed in samples at high temperature.

\section{Conclusion}
This investigation has used DFT to investigate the thermodynamics of the precipitation of Zr hydrides over a range of temperatures, compositions and starting solid solution concentrations. The use of statistically significant numbers of randomly generated configurations has been coupled with SQS cells to ensure that disordered cells are modelled accurately. This investigation has lead to the following conclusions:

\begin{itemize}
\item H favours a bimodal distribution within the Zr lattice. At low concentration, it prefers to maintain a dilute, non-clustered configuration, with a high energy barrier to hydride formation. As more H is absorbed by the Zr this barrier is overcome and hydride precipitation become energetically favourable.

\item The predicted concentration of the H solutions required to initiate precipitation is greater than observed experimentally, suggesting their may be additional mechanisms needed to enhance local H concentration to drive precipitation. Stress may play a part in this.

\item The calculation of latent enthalpies alone are insufficient to fully describe this system. Vibrational entropy, configurational entropy and sensible enthalpy are important for dealing with phase stabilities of precipitates and solid solutions.

\item Sensible enthalpies drive the reaction towards precipitation and are particularly significant for the $\zeta$-hydride.

\item Configurational entropy drives the system towards solution. They are particularly significant when contemplating the difference between HCP and FCC based hydrides.

\item Vibrational entropy and thus temperature drives the system towards solution.

\item Generally, the $\gamma$ phase is the most stable, suggesting other mechanisms, (such as precipitate interface lattice strain) may be responsible for the observed presence of $\delta$ hydrides.

\end{itemize}

\section*{Acknowledgements}

S. C. Lumley and M. R. Wenman acknowledge financial support from the MoD from a UDS grant.  M. R. Wenman additionally  acknowledges support from EDF Energy through an industrial Fellowship award. S. T. Murphy acknowledges financial support through an EPSRC grant (EP/I003320/1). All authors acknowledge the use of Imperial College's high performance computing centre.
\pagebreak

\appendix
\setcounter{table}{0}
\renewcommand{\thetable}{\Roman{table}}

\begin{table}[H]
\begin{center}
\begin{tabular}{cccc}
\hline
		& Property                           & Predicted Value& Literature  \tabularnewline
\hline
\hline
Zr (HCP)& a ($\angstrom$)                          &  3.22           & 3.23 \cite{Goldak1966} \tabularnewline
		& c ($\angstrom$)                          &  5.20           & 5.145 \cite{Goldak1966} \tabularnewline
		& $E_{vaporisation}$ (eV/atom)       &  6.20 (6.24)          & 6.24 \cite{CRC1999} \tabularnewline
\hline
Zr (FCC)& a ($\angstrom$)                          &  6.41        & - \tabularnewline
		& $E_{FCC}^{f}-E_{HCP}^{f}$ (eV/atom)&  0.04           & -  \tabularnewline
\hline
H (Gas) & Bond Length ($\angstrom$)                &  0.75              & 0.74 \cite{CRC1999} \tabularnewline
		& $E_{disassociation}$ (eV/atom)     & 4.53               & 4.52 \cite{CRC1999}\tabularnewline
		& Vibrational Frequency (cm$^{-1}$)  &  4328.42              & 4401.21 \cite{PhysicalChemistry1997} \tabularnewline
\hline
\end{tabular}
\caption{Crystallographic and thermodynamic properties of Zr and H. The bracketed value for $\alpha$ Zr includes the zero point energy component of atomic vibrations.} 
\label{tab:elements}
\end{center}
\end{table}

\begin{table}[H]
\begin{center}
\begin{tabular}{ccccc}
\hline
Structure                             & at\% H& Property       & Predicted Value\tabularnewline
\hline
\hline
Dilute HCP solution (tetrahedral site)& 1     & a ($\angstrom$)             &    2.24\tabularnewline
									&       & c ($\angstrom$)             &   5.19  \tabularnewline
									&       & $E_{solution}$ (eV) &     -0.60   \tabularnewline
\hline 							
Dilute HCP solution (octahedral site) & 1     & a ($\angstrom$)             &    6.23    \tabularnewline
									&       & c  ($\angstrom$)            &     5.17   \tabularnewline
									&       & $E_{solution}$ (eV) &      -0.56   \tabularnewline
\hline 									
Dilute FCC solution (tetrahedral site)& 11    & a ($\angstrom$)             &      3.23   \tabularnewline
									&       & $E_{solution}$ (eV) &        -0.19   \tabularnewline
\hline 									
Dilute FCC solution (octahedral site) & 11    & a ($\angstrom$)             &       3.22   \tabularnewline
									&       & $E_{solution}$ (eV)&       -0.43   \tabularnewline
\hline 									
$\zeta$ Hydride                       & 25    & a ($\angstrom$)             &      3.25    \tabularnewline
									&       & c ($\angstrom$)             &         10.78   \tabularnewline
									&       & $E_{formation}$ (eV/atom)&        -0.23  \tabularnewline
\hline 									
$\gamma$ Hydride                      & 50    & a ($\angstrom$)             &      4.57    \tabularnewline
									&       & c ($\angstrom$)             &        5.01   \tabularnewline
									&       & $E_{formation}$ (eV/atom)&        -0.44   \tabularnewline
\hline 									
$\delta$ Hydride [100]             & 60    & a ($\angstrom$)             &      4.75    \tabularnewline
									&       & c ($\angstrom$)             &        4.83    \tabularnewline
									&       & $E_{formation}$ (eV/atom)&        -0.50   \tabularnewline
\hline 									
$\delta$ Hydride [110]             & 60    & a ($\angstrom$)             &      4.69   \tabularnewline
									&       & c ($\angstrom$)             &        4.94    \tabularnewline
									&       & $E_{formation}$ (eV/atom)&        -0.51    \tabularnewline
\hline 									
$\delta$ Hydride [111]             & 60    & a ($\angstrom$)             &      4.77    \tabularnewline
									&       & $E_{formation}$ (eV/atom)&        -0.52   \tabularnewline
\hline 									
$\delta$ Hydride (SQS Average)             & 60    & a ($\angstrom$)             &      4.78    \tabularnewline
									&       & $E_{formation}$ (eV/atom)&        -0.53   \tabularnewline	
\hline 								
$\varepsilon$ Hydride                    & 66    & a  ($\angstrom$)            &      5.00   \tabularnewline
									&       & c  ($\angstrom$)            &        4.42   \tabularnewline
									&       & $E_{formation}$ (eV/atom)&        -0.59   \tabularnewline
\hline
\end{tabular}
\caption{Enthalpies and lattice parameters of ordered crystals.}
\label{tab:ordered}
\end{center}
\end{table}

\begin{table}[H]
\begin{center}
\begin{tabular}{cccccccc}
\hline
			      & $N$ & $\sum H$ & $R_{l}$& $R_{u}$ & $\bar{x}_{stoich}$& $\sigma_{stoich}$ \tabularnewline
\hline
\hline
Solid Solution & 50                   & 800                           & 0.13                  & 0.50                      & 0.19              & 0.08 \tabularnewline
Zeta Hydride      & 50                   & 800                           & 0.44                  & 0.88                      & 0.57              & 0.09 \tabularnewline
Randomised Gamma  & 15                   & 480                           & 0.63                  & 1.25                      & 1.00              & 0.16 \tabularnewline
Gamma Hydride     & 50                   & 1600                          & 0.75                  & 1.38                      & 1.03              & 0.18 \tabularnewline
Delta Hydride     & 50                   & 1600                          & 1.25                  & 1.88                      & 1.66              & 0.16 \tabularnewline
SQS Delta Hydrides& 3                    & 192                           & 1.50                  & 1.75                      & 1.63              & 0.10 \tabularnewline
\hline
\end{tabular}

\caption{Statistical values collected from the different data sets used in this work. $N$ is the size of the set, $\sum H$ is the total number of H sites, $R_{l}$ and $R_{u}$ are the lower and upper ranges of the stoichiometry, $\bar{x}_{stoich}$ is the arithmetic mean of the stoichiometry and $\sigma_{stoich}$ is the standard deviation from the mean.}
\label{tab:Stats}
\end{center}
\end{table}

\setcounter{figure}{0}
\renewcommand{\thefigure}{\arabic{figure}}

\begin{figure}[H]
\begin{center}
\includegraphics[scale=0.38]{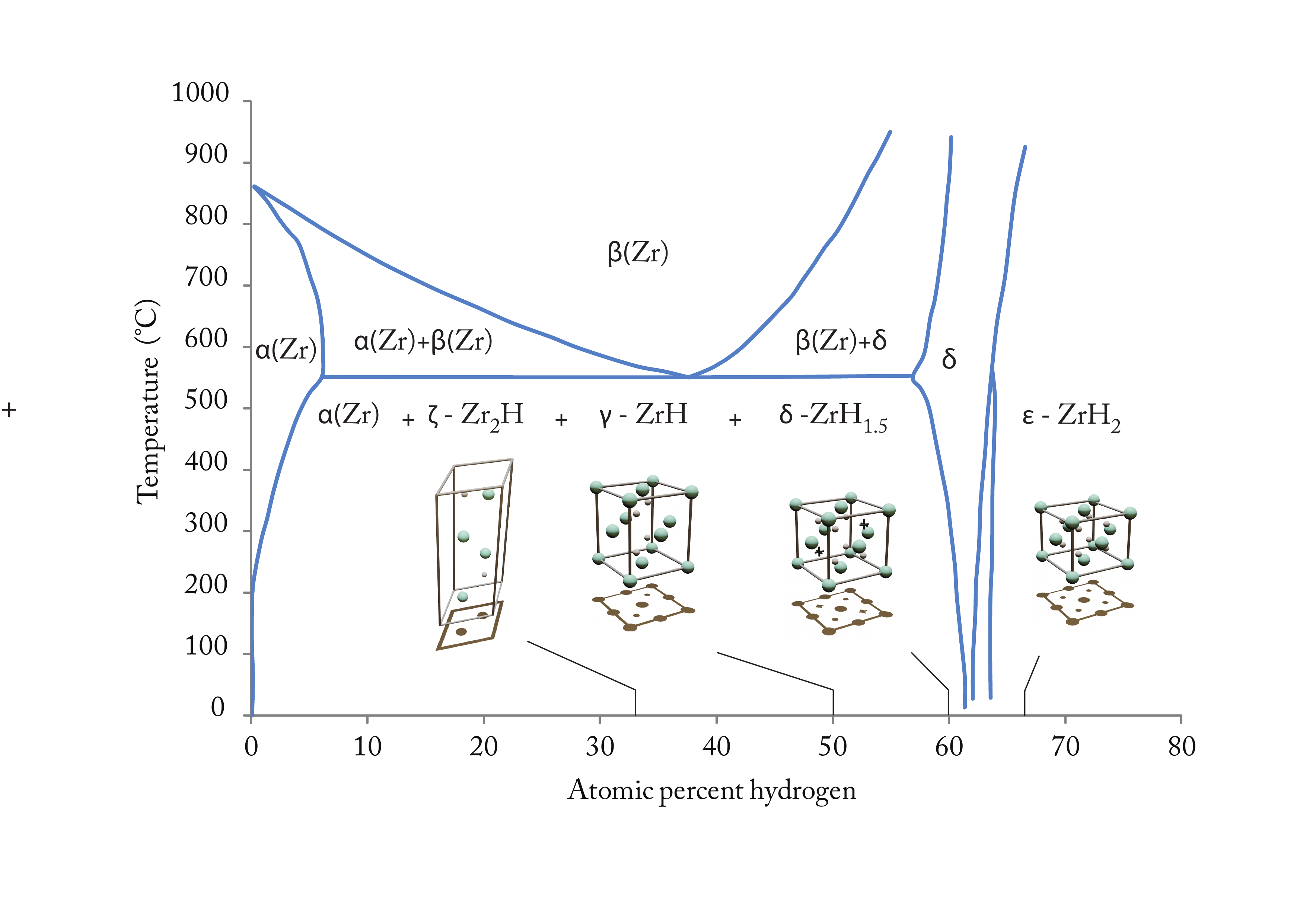}
\caption{Structures and formulae of different hydrides found in the Zr-H system. The Zr and H atoms are represented by the larger and smaller spheres respectively. The $\delta$ Hydride shown here is as commonly simulated, while real $\delta$-phase hydrides have a formula of ZrH$_{1.6}$ and have H atoms arranged randomly across all tetrahedral sites, including the black crosses shown in the diagram. The phase diagram has been reproduced from \cite{Okamoto2006b}.}
\label{fig:Hydrides}
\end{center}
\end{figure}

\begin{figure}[H]
\begin{center}
\includegraphics[scale=0.38]{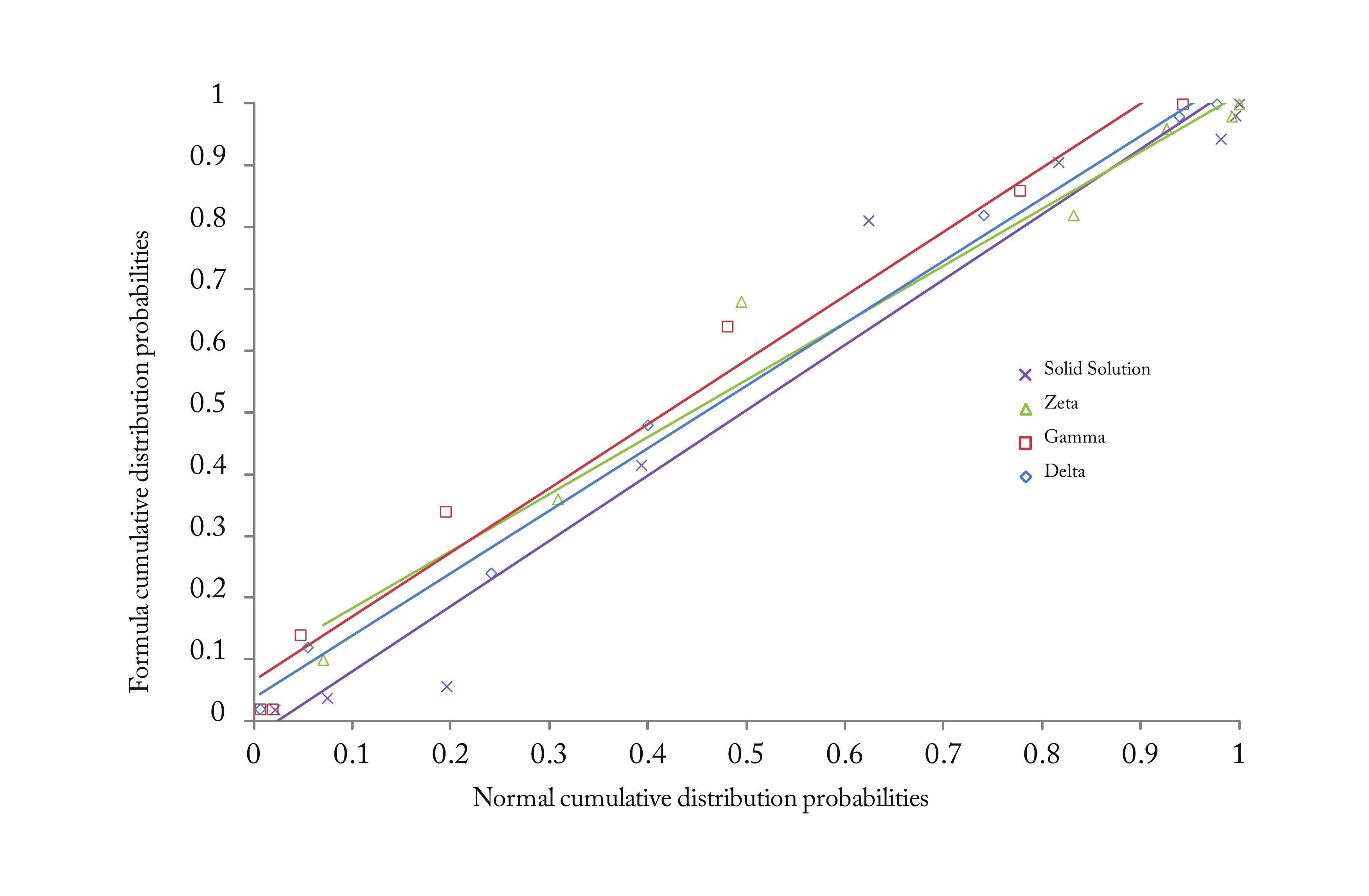}
\caption{Normality test plot for the structures generated in this study. The results are generated from a frequency distribution analysis of values for $x$ in the sample of  ZrH$_{x}$ structures.}
\label{fig:Statform}
\end{center}
\end{figure}

\begin{figure}[H]
\begin{center}
\includegraphics[scale=0.38]{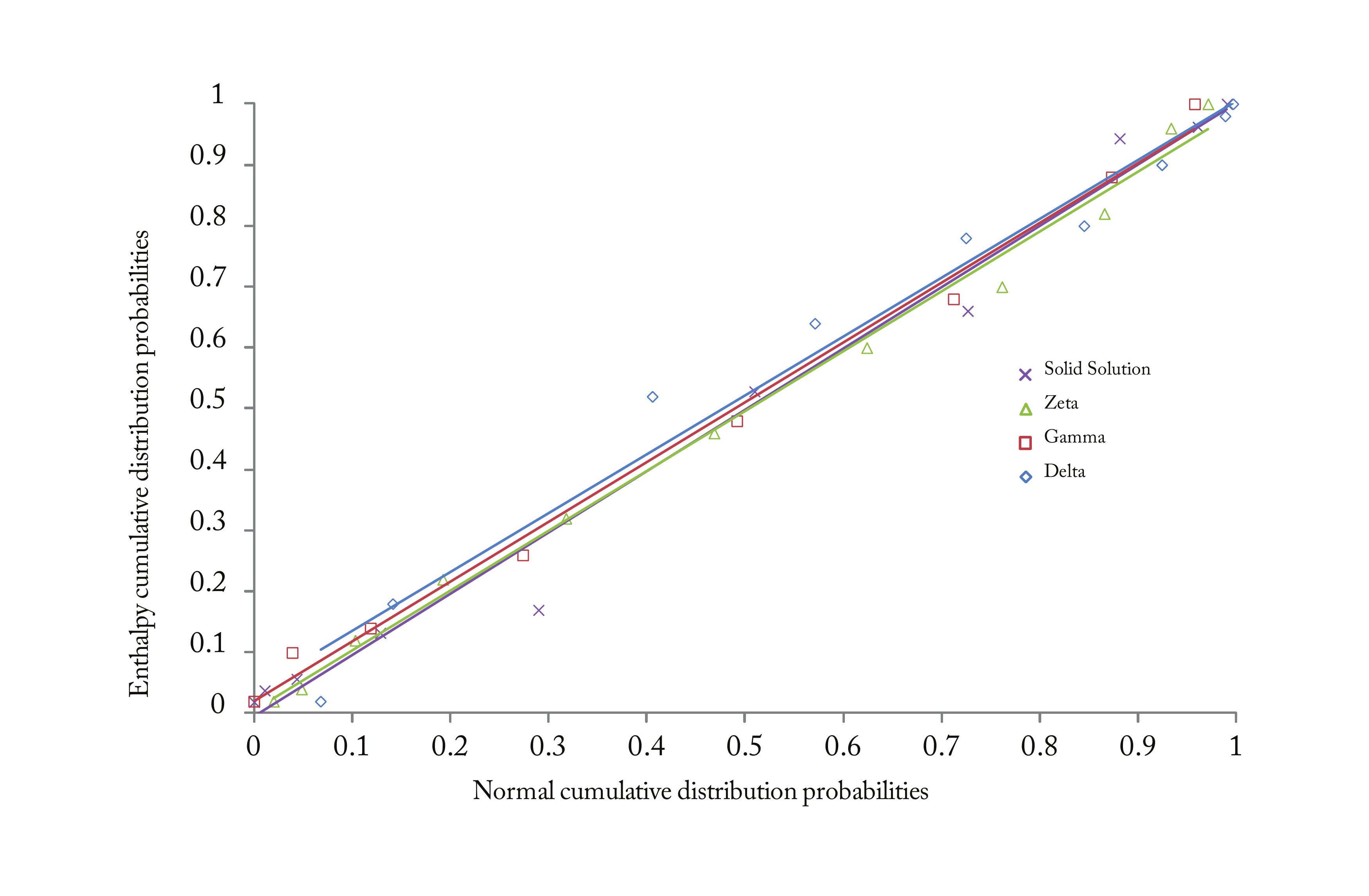}
\caption{Normality test plot for the structures generated in this study. The results are generated from a frequency distribution analysis of formation enthalpy values in the sample of structures.}
\label{fig:Statenth}
\end{center}
\end{figure}

\begin{figure}[H]
\begin{center}
\includegraphics[scale=0.38]{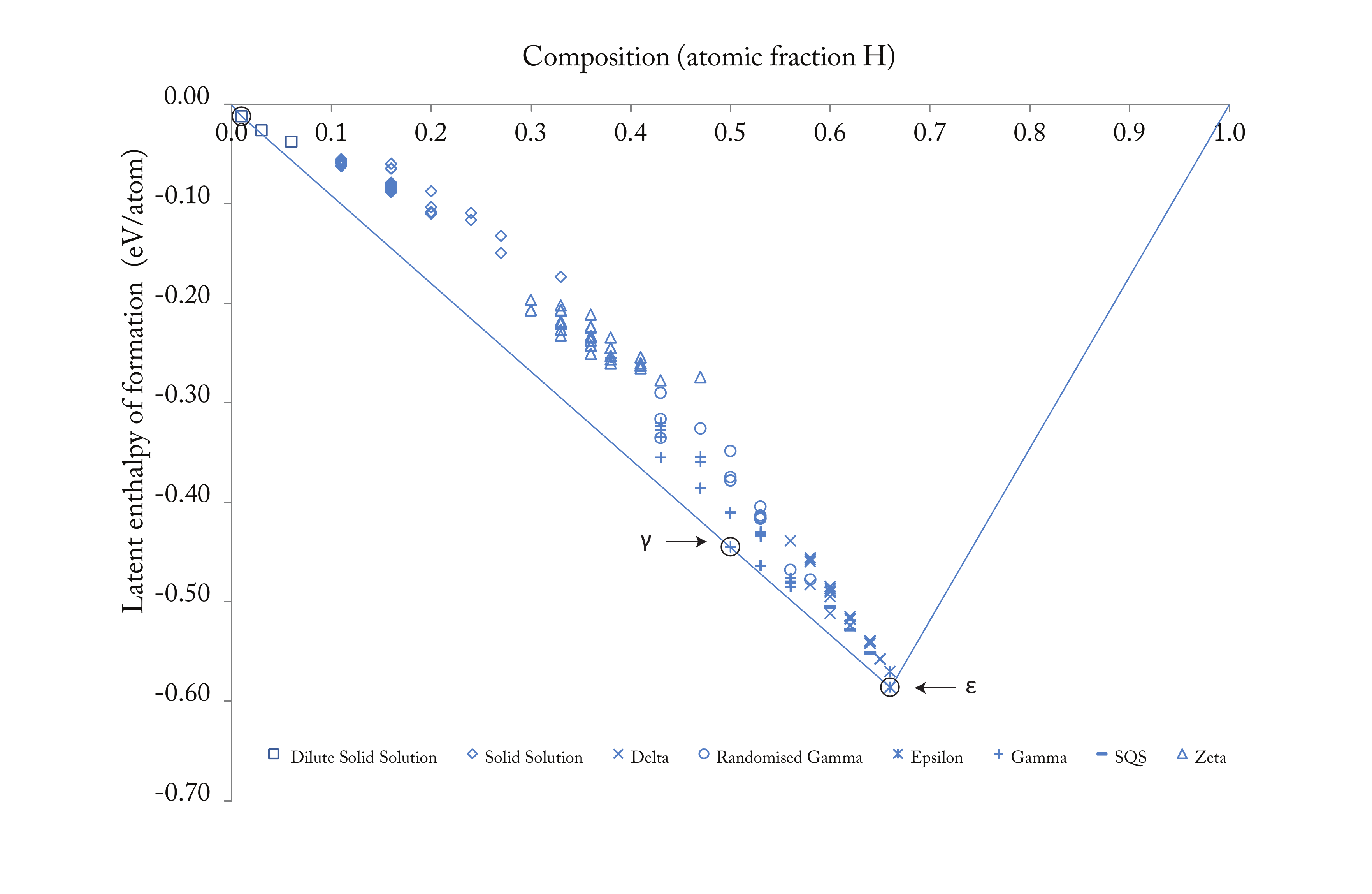}
\caption{The normalised latent enthalpy of formation from pure elements for all structures studied in this work. The shape of the marker indicates which simulation series it belongs to. Circled points rest on the convex hull.}
\label{fig:FormationEnthaply}
\end{center}
\end{figure}

\begin{figure}[H]
\begin{center}
\includegraphics[scale=0.38]{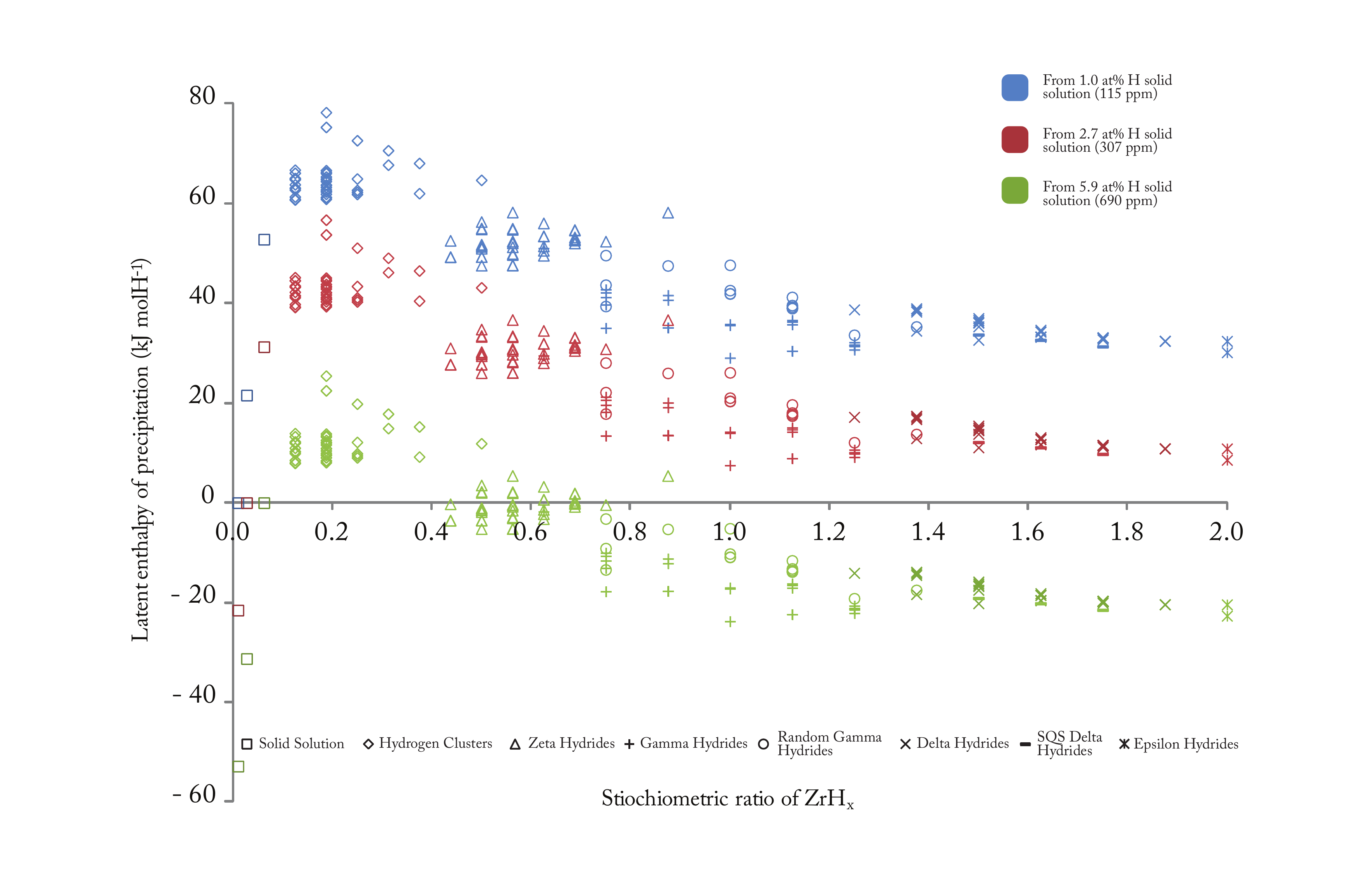}
\caption{The normalised latent enthalpy for all structures studied in this work. The shape of the marker indicates which simulation series it belongs to, and the colour indicates the concentration of the initial solid solution. These numbers are normalised with respect to the number H atoms precipitated.}
\label{fig:LatentEnthaply}
\end{center}
\end{figure}

\begin{figure}[H]
\begin{center}
\includegraphics[scale=0.38]{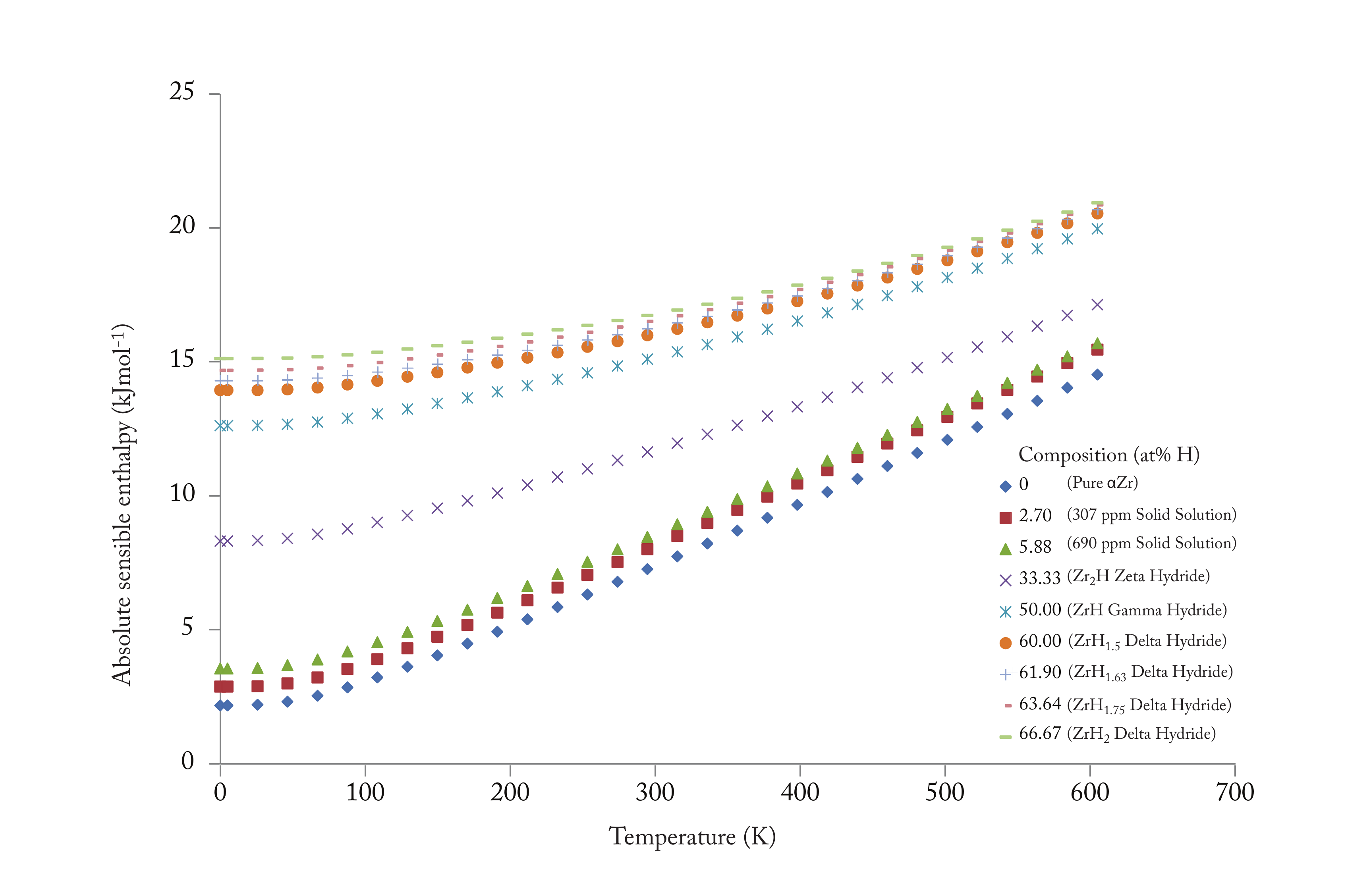}
\caption{The absolute sensible enthalpy plotted against temperature for a variety of reference simulations. The temperature-composition-enthalpy surface resultant from this data is used to interpolate sensible enthalpies for further calculations.}
\label{fig:AbsSensibleEnthalpy}
\end{center}
\end{figure}

\begin{figure}[H]
\begin{center}
\includegraphics[scale=0.38]{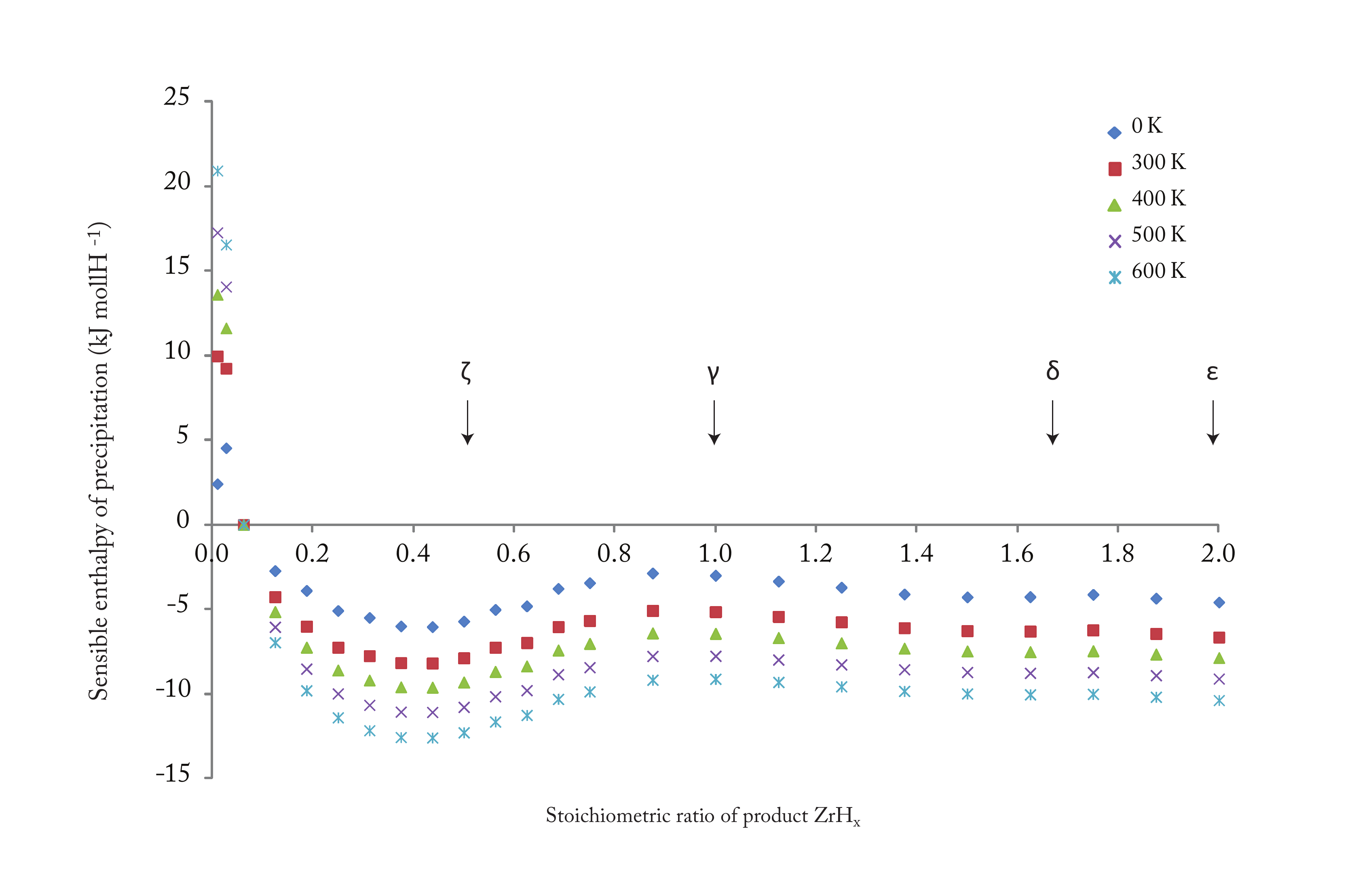}
\caption{The sensible enthalpy change during precipitation as a function of stoichiometry. These numbers are normalised with respect to the number of H atoms precipitated. }
\label{fig:SensibleEnthalpy}
\end{center}
\end{figure}

\begin{figure}[H]
\begin{center}
\includegraphics[scale=0.38]{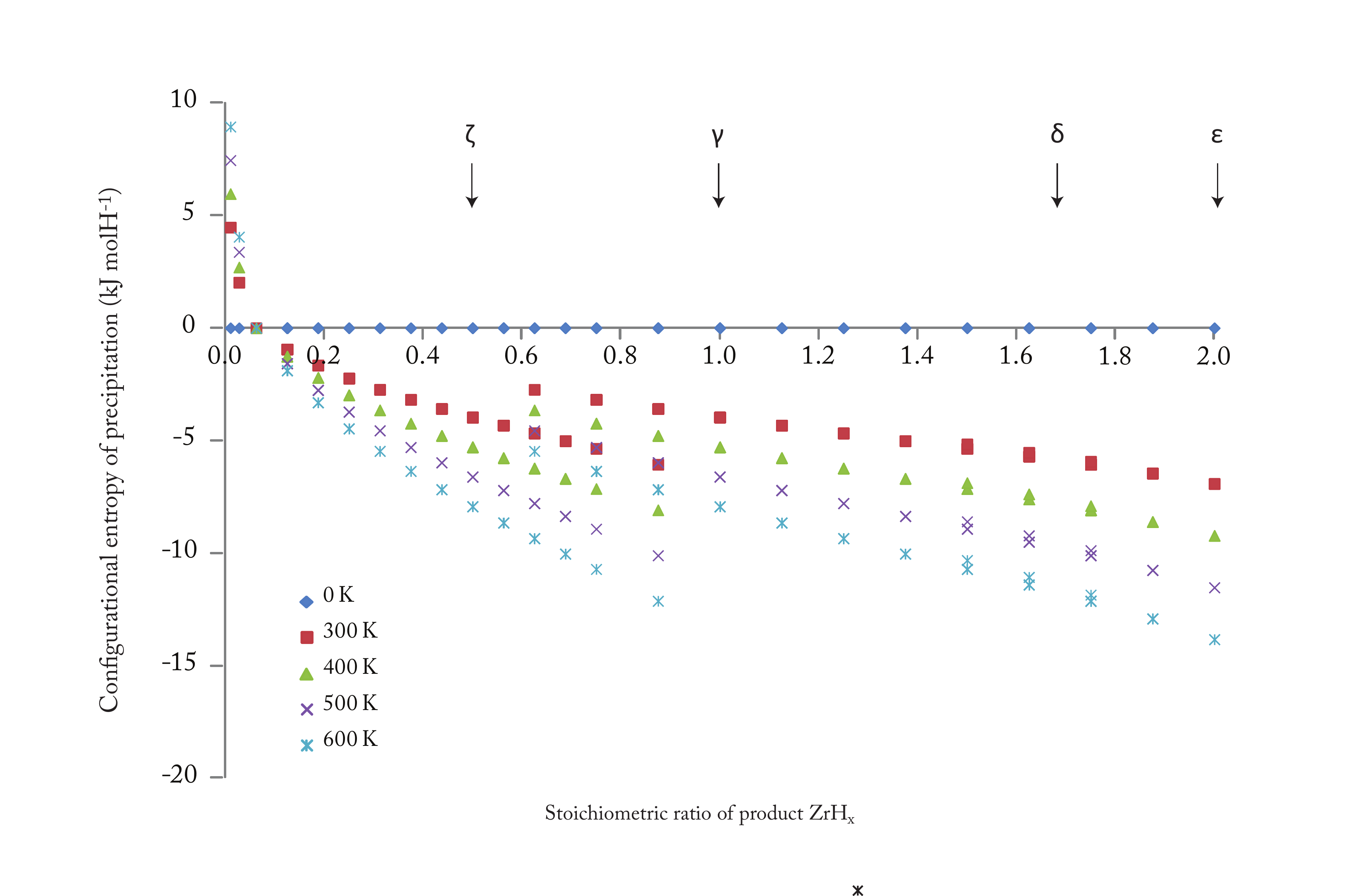}
\caption{The configurational entropy change during precipitation as a function of stoichiometry. These numbers are normalised with respect to the number of H atoms precipitated. It should be noted that these are presented as a T$\Delta$S product.}
\label{fig:ConfigEntropy}
\end{center}
\end{figure}

\begin{figure}[H]
\begin{center}
\includegraphics[scale=0.38]{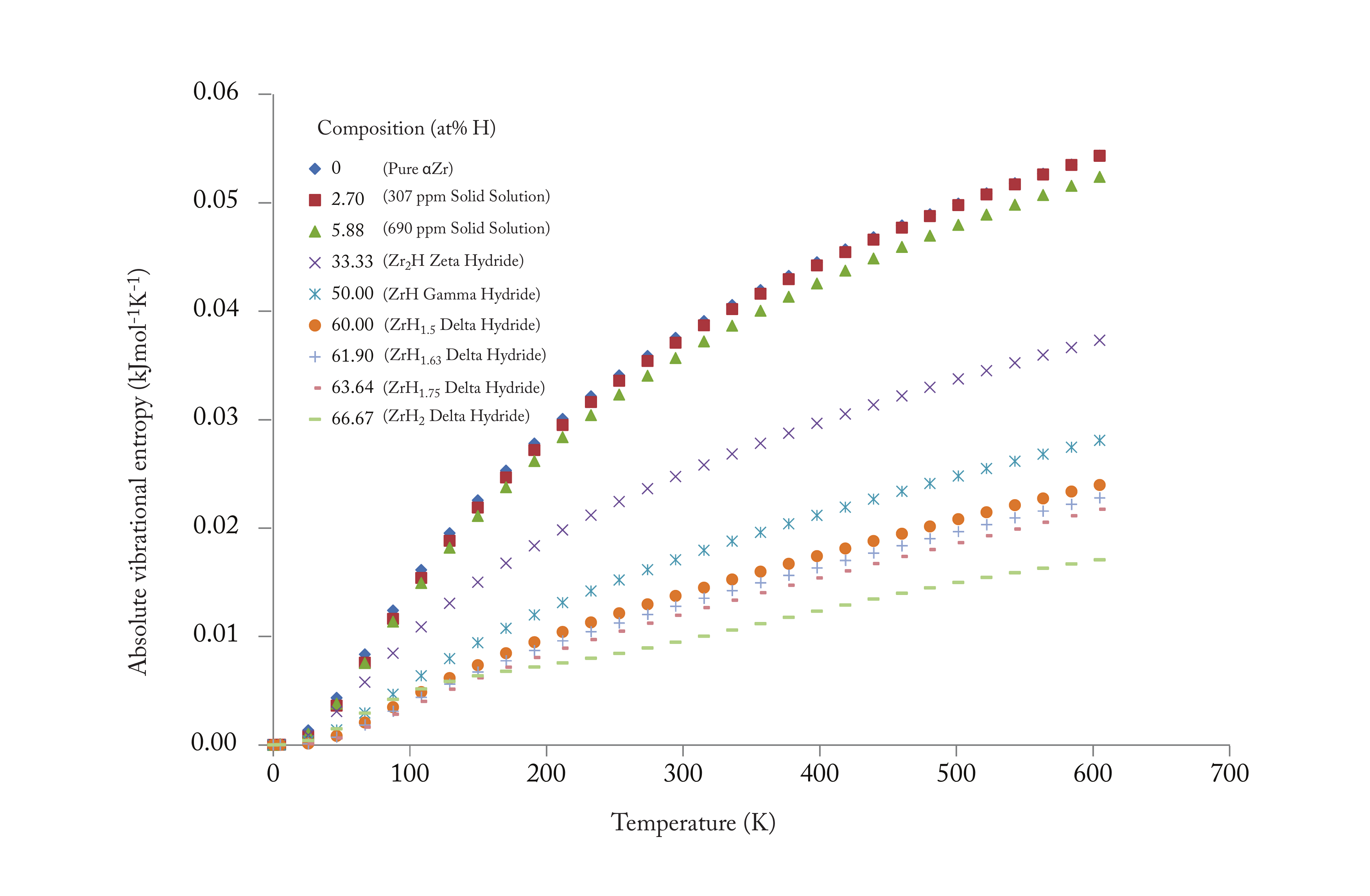}
\caption{The absolute vibrational entropy plotted against temperature for a variety of reference cells. The temperature-composition-entropy surface resultant from this data is used to interpolate vibrational entropies for further calculations.}
\label{fig:AbsVibrationalEntropy}
\end{center}
\end{figure}

\begin{figure}[H]
\begin{center}
\includegraphics[scale=0.38]{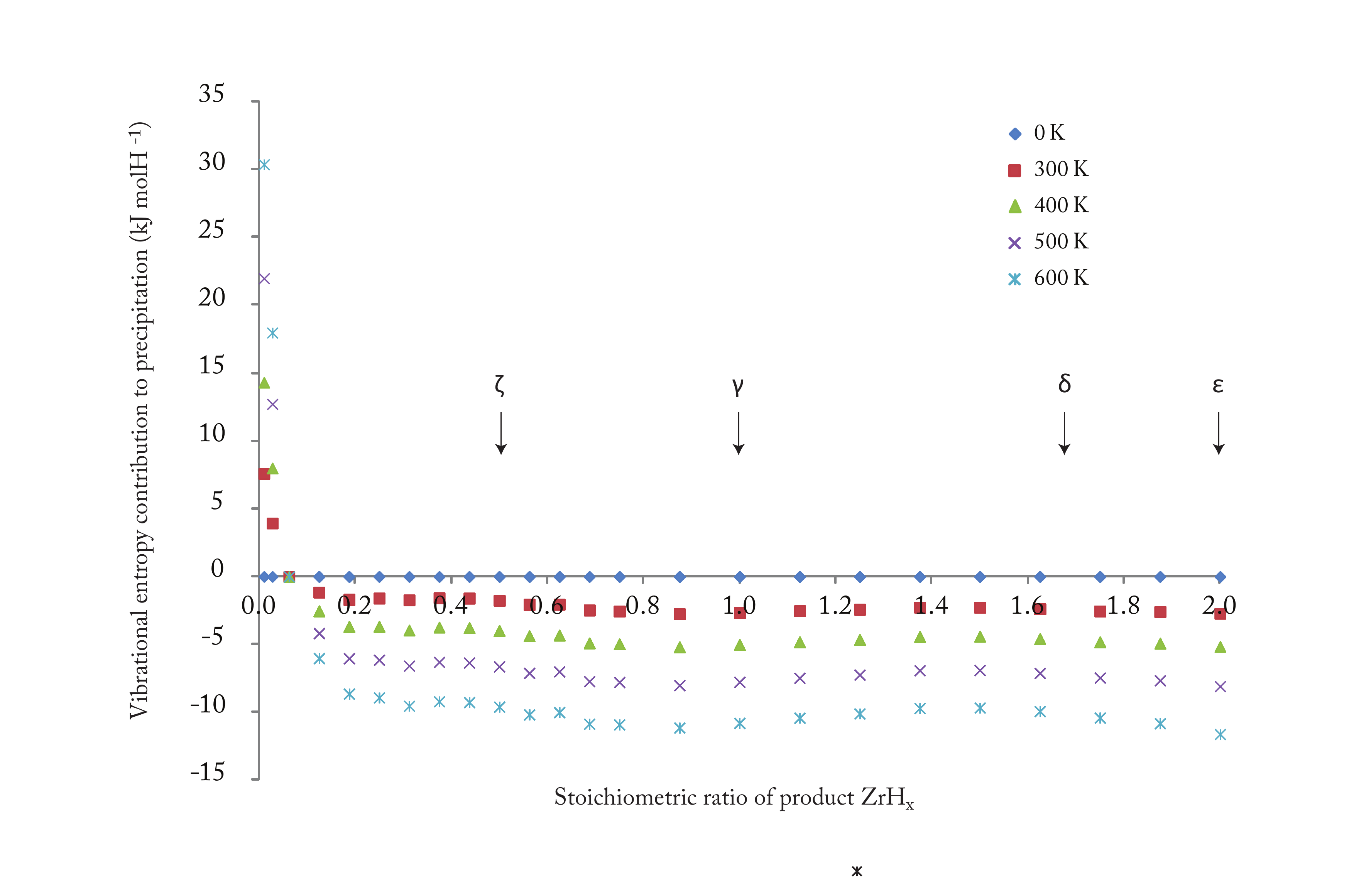}
\caption{The vibrational entropy change during precipitation as a function of stoichiometry. These numbers are normalised with respect to the number of H atoms precipitated. It should be noted that these are presented as a T$\Delta$S product.}
\label{fig:VibrationalEntropy}
\end{center}
\end{figure}

\pagebreak

\begin{figure}[P]
     \centering
     \subfigure[\label{fig:FreeEnergy1}Precipitation from from a 1.0 at\%H solid solution.]{
          \includegraphics[scale=0.3]{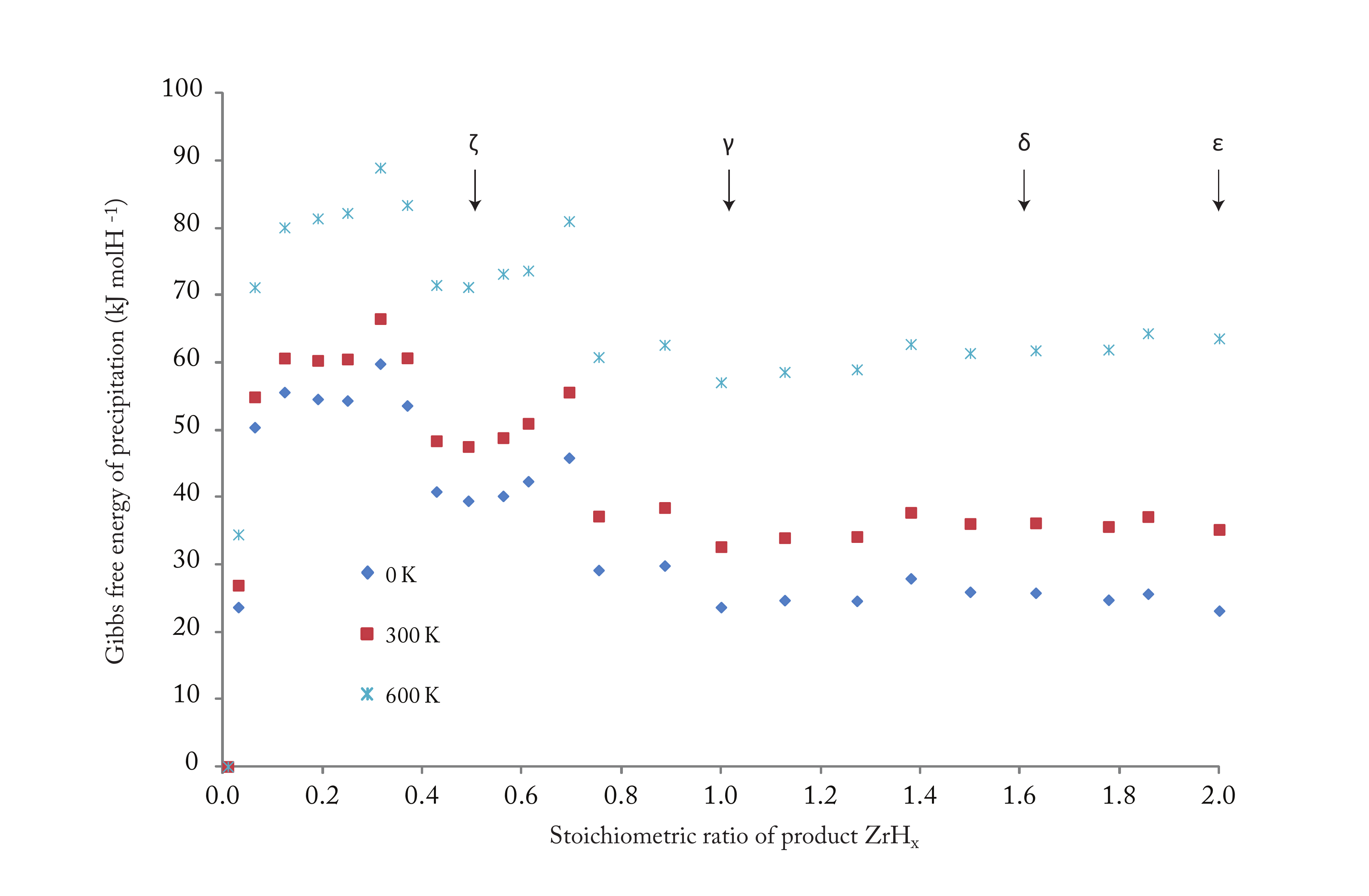}} \\
     \subfigure[\label{fig:FreeEnergy2}Precipitation from from a 2.7 at\%H solid solution.]{
          \includegraphics[scale=0.3]{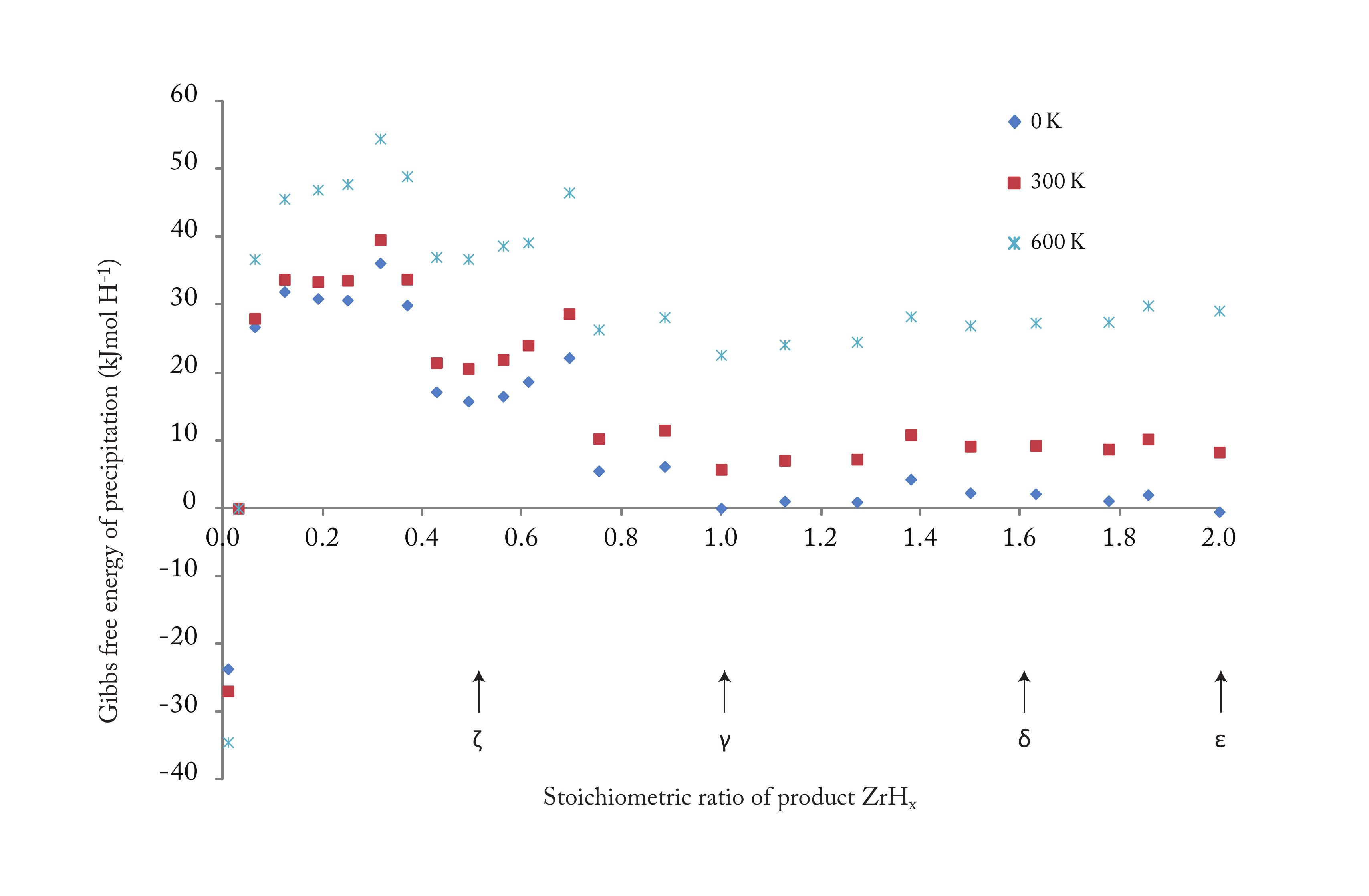}}
     \subfigure[\label{fig:FreeEnergy3}Precipitation from from a 5.9 at\%H solid solution.]{
          \includegraphics[scale=0.3]{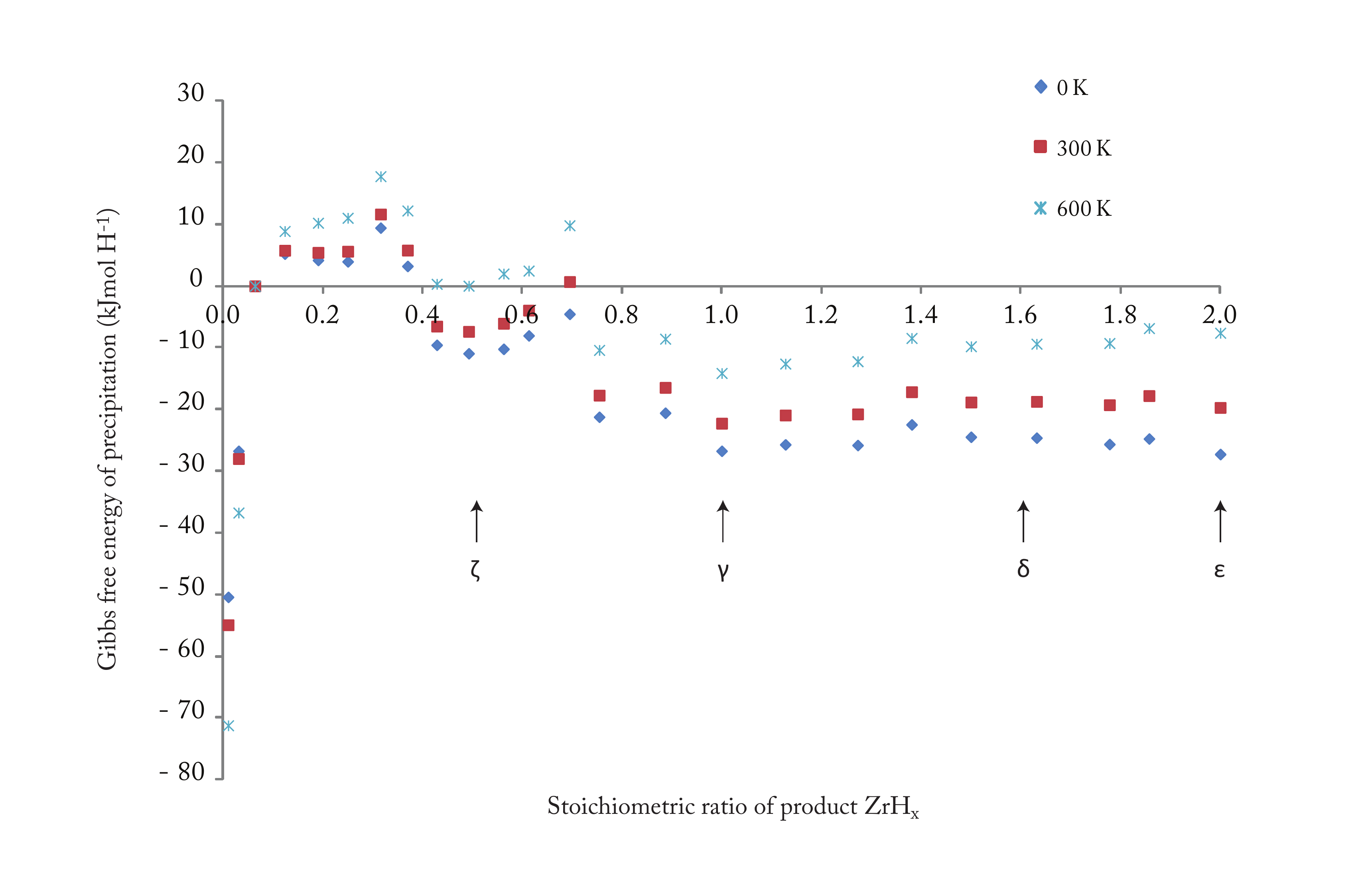}}
     \caption{\label{fig:FreeEnergies}The final Gibbs free energy of precipitation, as a function of stoichiometry, from different starting concentrations. These numbers are normalised with respect to the number of H atoms precipitated.}
\end{figure}

\bibliographystyle{elsarticle-num}

\end{document}